\documentclass{llncs}
\usepackage{amsmath}
\usepackage{bm}
\usepackage{multirow}
\usepackage{graphicx}
\usepackage{epstopdf}
\usepackage{epsfig}
\usepackage{makeidx}

\begin{document}

\title{Summarizing Reviews with Variable-length Syntactic Patterns and Topic Models}
\author{Trung Nguyen\inst{1} \and Alice Oh\inst{1}}
\institute{Computer Science Department, KAIST, Daejeon 305-701, South Korea}
\maketitle

\begin{abstract}
We present a novel summarization framework for reviews of products and services by selecting informative and concise text segments from the reviews. Our method consists of two major steps. First, we identify five frequently occurring variable-length syntactic patterns and use them to extract candidate segments. Then we use the output of a joint generative sentiment topic model to filter out the non-informative segments. We verify the proposed method with quantitative and qualitative experiments. In a quantitative study, our approach outperforms previous methods in producing informative segments and summaries that capture aspects of products and services as expressed in the user-generated pros and cons lists. Our user study with ninety users resonates with this result: individual segments extracted and
filtered by our method are rated as more useful by users compared to previous approaches by users.
\end{abstract}

\section{Introduction}
Online reviews of products and services are an important source of knowledge for people to make their purchasing decisions. They contain a wealth of information on various product/service aspects from diverse perspectives  of consumers. However, it is a challenge for stakeholders to retrieve useful information from the enormous pool of reviews.  Many automatic systems were built to address this challenge including generating aspect-based sentiment summarization of reviews  \cite{Blair:google,Hu:2004,Popescu:2005} and comparing and ranking products with regard to their aspects \cite{Liu:nlphandbook}.  In this study we focus on the problem of review summarization, which takes as input a set of user reviews for a specific product or service entity  and produces a set of representative text excerpts from the reviews. 

Most work on summarization so far used sentence as the unit of summary. However, we do not need a complete
sentence to understand its main communicative point. Consider the following sentence from review of a coffee maker:
`My mum bought me this one, and I have to say it makes really awful tasing coffee'. To a buyer looking for an opinion
about the coffee maker, only the part `makes really awfultasing coffee' is helpful. Being able to extract such short
and meaningful segments from lengthy sentences can bring significant utilities to users. It reduces their reading load as well as presents more readable summaries on devices with limited screen size such as smart phones.

This motivates our main research question of how to extract concise and informative text from reviews of products and services that can be used for summarization. Previous work has ignored the differences in product and service reviews, which is questionable. To the best of our knowledge, this is the first work that studies and compares summarization for the two domains in details. We propose to to extract text segments that match against pre-defined syntactic patterns that occur frequently in reviews of both products and services. However, the extracted segments should be subjected to some selection or filtering procedure as not all matching candidates are likely to contain rich information. Our proposed selection mechanism is based on the observation that segments containing users' opinions and evaluations about product and service aspects carry valuable information. This motivates the use of output of joint sentiment topic models to discriminate between desirable and non-desirable text segments. Since joint sentiment topic models capture sentiments that are highly associative with aspects, they are well suited for selecting informative segments from the pool of extracted candidates.

The major contributions of our work are as follows.
\begin{enumerate}
\item A new joint sentiment-topic model that automatically
 learns polarities of sentiment lexicons from reviews.
\item Identification of five frequently occuring syntactic patterns for extracting concise segments from reviews of both products and services.
\item Demonstration of the effective application of topic models to select informative variable-length
segments for review summarization.
\item Production of summaries that recall important information from review entities' characteristics.
\end{enumerate}

The rest of the paper is structured as follows. We begin with the related literature in review summarization and joint
sentiment topic models in Sect. 2. Next we describe our extension to a topic model and its improvements over previous models in Sect. 3. We then introduce our proposed extraction patterns and procedures for segment selection in Sect. 4. We present our experiments and evaluation in Sect. 5 and 6 and conclude in Sect. 7.

\section{Related work}
We first look at how text excerpts are extracted from reviews in the existing literature. Previous studies mainly generated aspect-based summary for products and services by aggregating subjective text excerpts related to each aspect. Different forms of the excerpts include sentence \cite{Hu:2004}, concise phrase composing of a modifier and a header term \cite{Lu:shortcomments}, adjective-noun pair extracted based on POS tagging and the term-frequency of the pair \cite{Yatani:spotlight}, and phrase generated by rules \cite{Liu:paraphrase}. Some limitations of these previous work are i) they only worked with the simplistic adjective-noun pairs or specific form of reviews such as short comments, and ii) experiments were carried out with reviews of services only. Our approach to extract text segments by matching variable-length linguistic patterns overcome these shortcomings and can generalize well for free-text reviews of both products and services. 

Various methods for selecting informative text fragments were applied in previous research, such as matching against pre-defined or frequently occurring aspects \cite{Blair:google,Hu:2004}, ranking frequency \cite{Yatani:spotlight}, and topic models \cite{Mei:tsm,Titov:mas,Xu:2011}. We are interested in the application of joint sentiment topic models as they can infer sentiment words that are closely associative with an aspect. This is an important property of polarity of sentiment words as pointed out in \cite{Fahrni:oldwine,Lin:jst,Liu:nlphandbook,Pang:thumbsup}, and recently several joint topic models  have been proposed to unify the treatment of sentiment and topic (aspect) \cite{Jo:asum,Lin:jst,Mei:tsm,Titov:mga}. Applications of these models have been limited to sentiment classification for reviews, but we hypothesize that they can also be helpful in summarization. We focus our next discussion on previous joint models in comparison to our proposed model.

One of the earliest work is the Topic-Sentiment Model (TSM) \cite{Mei:tsm}, which generates a word either from a topic or one
 of the two additional subtopics -- sentiments, but it fails to account for the intimate interplay between a topic/aspect and a sentiment. TSM is based on pLSI whereas more recent work (\cite{Jo:asum,Lin:jst,Titov:mas}) uses or extends Latent Dirichlet Allocation (LDA) \cite{Blei:lda}. In the Multi-Aspect Sentiment (MAS) model \cite{Titov:mas}, customer ratings are incorporated as signals to guide the formation  of pre-defined aspects, which can then be used to extract sentences from reviews that are related to each aspect.
  In the Joint Sentiment/Topic (JST) model \cite{Lin:jst}, and the Aspect and Sentiment Unification Model (ASUM) \cite{Jo:asum},  each word is assumed to be generated from a distribution jointly defined by a topic and a sentiment (either positive or negative).  As a result, JST and ASUM learn words that are commonly associated with an aspect although the models are incapable of distinguishing between  sentiment and non-sentiment lexicons.
We propose a new model that leverages syntactic information to identify sentiment lexicons  and automatically learn their polarities from the co-occurrences of words in a sentence. This allows the model to bootstrap using  a minimum set of sentiment seed words, thereby alleviating the need for information that is expensive to obtain
 such as ratings of users for reviews \cite{Titov:mas} or large lists of sentiment lexicons \cite{Lin:jst}.

\section{A Topic Model for Learning Polarity of Sentiment Lexicons}
Our key modelling assumption for reviews is that a sentence expresses an opinion toward an aspect via its sentiment component. For example, in the sentence  `The service was excellent', only the word `excellent' carries the positive sentiment. This is not a new assumption  as adjectives and adverbs are commonly considered the main source of sentiment in a sentence in existing literature. Our model leverages on this type of knowledge to locate sentiment words in a sentence with relatively high confidence.
\begin{figure}
\centering
\includegraphics[scale=0.2]{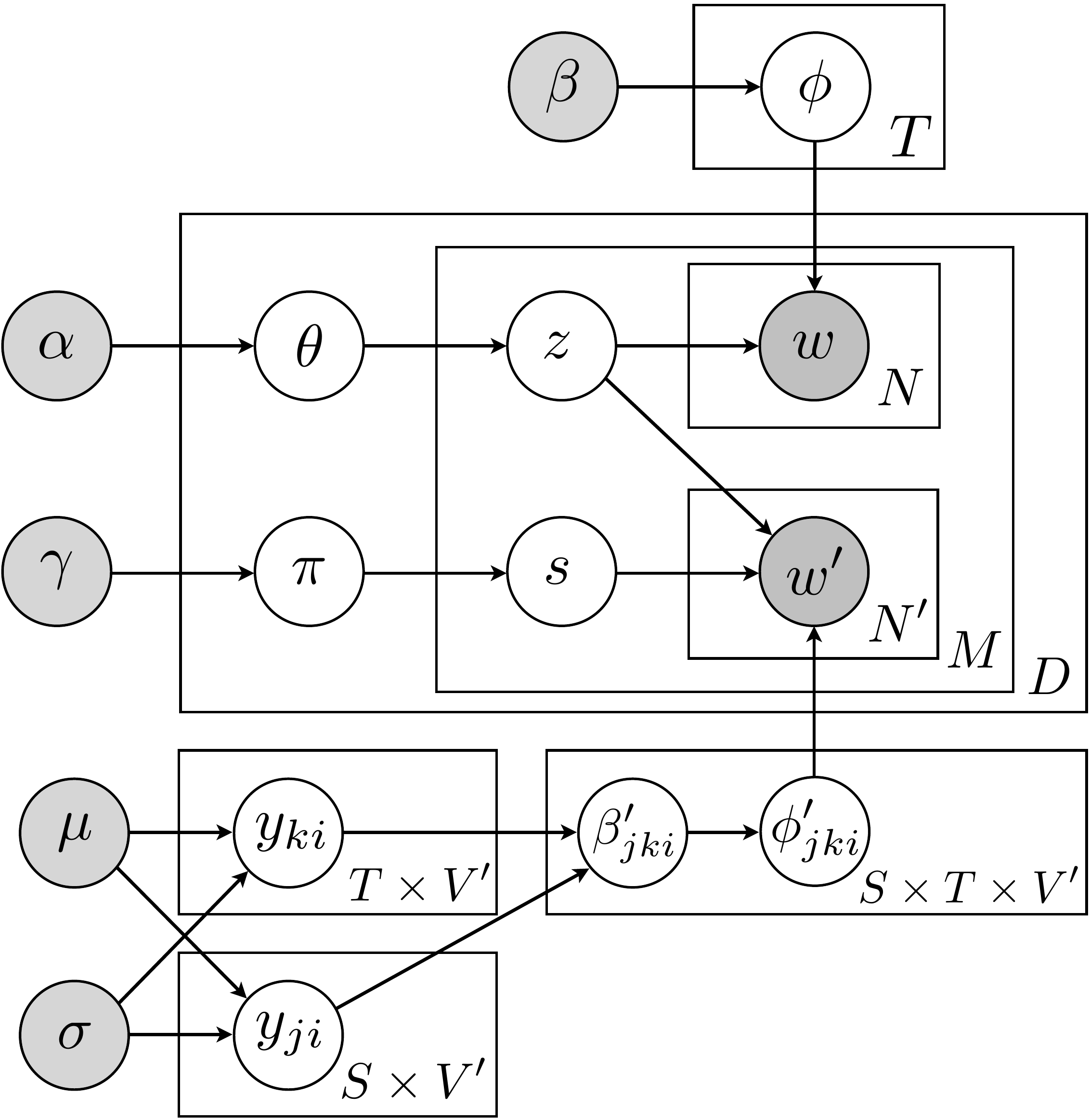}
\caption{Graphical representation of the model.}
\label{fig:model}
\end{figure}
 
\subsection{Generative Process}
The formal generative process of our model for the graphical representation in Fig. ~\ref{fig:model} is as follows (see Table ~\ref{tab:notations} for the list of notations).
\begin{itemize}
\renewcommand{\labelitemii}{$\diamond$}
\renewcommand{\labelitemiii}{\bf --}
\item{For every aspect $k$, draw a distribution of non-sentiment words, $\bm{\phi_{k}} \sim \mathrm{Dir}(\beta),$ } and two distributions of sentiment words, $\bm{\phi'_{jk}} \sim \mathrm{Dir}(\mathbf{\beta'_{jk}})$,
 where $j = 0$ denotes positive polarity and $j=1$ denotes negative polarity.
\item{For each review $d$,
\begin{itemize}
\item{Draw a sentiment distribution $\bm{\pi}_{d} \sim \mathrm{Dir}(\gamma)$}
\item{Draw a topic distribution $\bm{\theta}_{d} \sim \mathrm{Dir}(\alpha)$}
\item{For each sentence $c$ in document $d$,
\begin{itemize}
\item{Choose a topic $z = k \sim \mathrm{Mult}(\bm{\theta}_{d})$ 
and a sentiment $s = j \sim \mathrm{Mult}(\bm{\pi}_{d})$}
\item{Choose words $w \sim \mathrm{Mult}(\bm{\phi}_{k})$ to discuss aspect $k$ and sentiment words $w' \sim \mathrm{Mult}(\bm{\phi'}_{jk})$
 to convey the sentiment $j$ toward $k$.}
\end{itemize}
}
\end{itemize}
}
\end{itemize}
Notice in the graphical model that the part of a sentence which emanates the sentiment is observed.  In our implementation, we treat all adjectives and adverbs as $w'$ and remaining words as $w$ in the generative procedure, but this is not a restriction imposed on the model. It is easy to incorporate prior knowledge about words that convey sentiment into the model. For example, we can instruct the model that words such as \textit{love, hate, enjoy, worth, disappoint} are sentiment words, even though they are not adjective nor adverb.

Our main extension deals with the word smoother $\bm{\beta'}$ for sentiment words. Each sentiment word $i$ is associated with a topic dependent smoothing coefficient
 $y_{ki}$ for topic $k$ and a sentiment dependent smoothing coefficient $y_{ji}$ for sentiment $j$.
 We then impose that
\begin{alignat}{3}
\beta'_{jki} &= \exp(y_{ki} + y_{ji}),
&\quad y_{ki} &\sim N(0, \sigma^{2}_{1}),
&\quad y_{ji} &\sim N(0, \sigma^{2}_{2}).
\end{alignat}
This modeling allows us to incorporate polarity of sentiment words as side information.
The polarity of sentiment lexicon $i$ in a corpus is represented by the values of $y_{ji}$; this is to assume that  
the polarity of $i$ is its intrinsic property as the corpus is about a specific domain \cite{Choi:2009}.
 The topic dependent smoother $y_{ki}$ is introduced to accommodate the different frequency of association between
 the sentiment word $i$ and different aspects. 
\begin{table}
\centering
\caption{List of notations used in the paper (senti = sentiment, dist. = distribution)}
\label{tab:notations}
\begin{tabular}{|llp{10cm}|} \hline
$d, c, w, w', k, j$&:& review, sentence, non-senti word, senti word, topic/aspect, sentiment \\
$T, S, V, V'$&:& number of topics, sentiments, non-senti words, senti words\\
$\bm{\pi}_d, \bm{\theta}_d$&:&sentiment distribution, topic distribution of the review $d$ \\
$\bm{\phi}_k, \bm{\phi'}_{jk}$&:&word dist. of topic $k$, senti word dist. of topic $k$ and senti $j$ \\
$y_{ji}$&:&polarity of senti word $i$ with sentiment $j$ \\
$y_{ki}$&:&smoother for dependency between topic $k$ and senti word $i$\\
$\beta'_{jki}$&:&word smoother for senti word $i$ with topic $k$ and senti $j$ \\
$\alpha,\beta,\gamma,\mu,\sigma$&:& hyperparameters \\
$n^{TW}_{ki}, n^{STW}_{jki}$&:& counts of  word $i$ being assigned topic $k$, senti word $i$ being assigned topic $k$ and senti $j$\\
$n^{DT}_{dk}, n^{DS}_{dj}$&:& counts of sentences in $d$ being assigned topic $k$, sentences in $d$ being assigned senti $j$\\
\hline
\end{tabular}
\end{table}
\subsection{Inference}
In order to perform inference we alternate between two procedures: sampling and maximum a posteriori.
The sampler assigns values for the latent variables: the topics and sentiments of sentences.
 Using a collapsed Gibbs sampler \cite{Griffiths:gibbs}, new values for the topic and sentiment of a sentence $c$ in document $d$
 are drawn from the conditional probability
\begin{eqnarray} \nonumber
p(z_{dc} = k, s_{dc} = j|\mathrm{rest}) \propto  
\frac{\prod_{i \in A(dc)} (n^{TW}_{\backslash ki} + \beta)}{\prod_{x = 0}^{\lvert A(dc)\rvert - 1}\sum_{i = 1}^{V}n_{\backslash ki}^{TW} + V\beta + x} \\
\frac{\prod_{i \in S(dc)} (n^{STW}_{\backslash jki} + \beta'_{jki})}{\prod_{x = 0}^{|S(dc)| - 1}\sum_{i = 1}^{V'}(n_{\backslash jki}^{STW} + \beta'_{jki}) + x} 
 (n^{DT}_{\backslash dk} + \alpha) (n^{DS}_{\backslash dj} + \gamma)  
\end{eqnarray}
where $S(dc)$ is the set of sentiment words in $c$ and  $A(dc)$ is the set of remaining words.  The '\textbackslash' notation means not counting the sentence being sampled.

We estimate the value for $\bm{\beta'}$ and $\bm{y}$ from a maximum a posteriori procedure,
 optimizing $\bm{\beta'}$ over $y$ and the assigned values of the latent variables. The negative log prior is 
\begin{equation}
-\log p(\beta') = S\sum_{k,i}{y_{ki}} + T\sum_{j,i}{y_{ji}} + \sum_{j,k,i}{\frac{(y_{ki} + y_{ji})^2}{2\sigma^2}} 
\end{equation}
where $\sigma^2 = \sigma_{1}^2 + \sigma_{2}^2$.
The collapsed negative log likelihood (dependent on sentiment words only) is 
\begin{equation}
L_{\beta'} = \sum_{j,k}{\left[ \log\Gamma({\bar n_{jk} + \bar \beta'_{jk}}) - \log\Gamma(\bar \beta'_{jki})\right]} 
+ \sum_{j,k,i}{\left[\log\Gamma(\beta'_{jki}) - \log\Gamma(n^{STW}_{jki} + \beta'_{jki}) \right]} 
\end{equation}
where $\bar n_{jk} = \sum_{i = 1}^{V'}{n^{STW}_{jki}}$, $\bar \beta'_{jk} = \sum_{i = 1}^{V'}{\beta'_{jki}}$, and $\Gamma$ is the Gamma function. We use the L-BFGS optimizer \cite{Liu:lbfgs} to minimize the objective function $L_{\beta'} - \log p(\beta')$ by taking its partial derivatives with respect to $y_{ki}$ and $y_{ji}$.

A sample from the Markov chain in the sampler can be used to estimate the distributions of interest. The approximate probabilities of sentiment $j$ in document $d$ ($\hat\pi_{dj}$), topic $k$ in document $d$ ($\hat\theta_{dk}$),  non-sentiment word $i$ in topic $k$ ($\hat\phi_{ki}$), and sentiment word $i$ in topic $k$ and sentiment $j$ ($\hat\phi'_{jki}$) are
\begin{alignat}{2}\nonumber
\hat\pi_{dj} &= \frac{n_{dj}^{DS} + \gamma}{\sum_{j' = 1}^{S}n_{dj'}^{DS} + S\gamma} \enspace,
&\quad
\hat\theta_{dk} &= \frac{n_{dk}^{DT} + \alpha}{\sum_{k' = 1}^{T}n_{dk'}^{DT} + T\alpha} \enspace, \\
\hat\phi_{ki} &= \frac{n_{ki}^{TW} + \beta}{\sum_{i' = 1}^{V}n_{ki'}^{TW} + V\beta} \enspace,
&\quad
\hat\phi'_{jki} &= \frac{n_{jki}^{STW} + \beta'_{jki}}{\sum_{i = 1}^{V'}(n_{jki'}^{STW} + \beta'_{jki'})} \enspace.
\end{alignat}

\subsection{Aspect and Sentiment Classification Using Output of the Model}
\label{sec:classifiers}
As stated in the introduction, we attempt to use the outputs of this model to improve the selection of informative segments for
 summarization. We define the topic classifier of an arbitrary segment of $n$ words $G = (w_{1}, w_{2},$ $\ldots, w_{n})$ as
\begin{equation}
\label{eq:segtopic}
\arg\max_{k} p(k|G) = \arg\max_{k} \sum_{i = w_{1}}^{w_{n}} (\log\hat\phi_{ki} + \sum_{j}\log\hat\phi'_{jki}) \enspace.
\end{equation}
To classify the sentiment of a segment G, we use the sentiment value $y_{ji}$ learned from the model. We define the polarity of G as
\begin{equation}
polarity(G) := \sum_{\text{sentiment word i}  \in G} polarity(i) = \sum_{\text{sentiment word i} \in G}y_{0i} - y_{1i}\enspace.
\end{equation}
G is classified as positive if $polarity(G) >= 0$ and as negative if $polarity(G) < 0$.

\section{Summarization Using Syntactic Patterns and Topic Models}
In this section we present our framework for variable-length segment-based summarization of reviews.
We first describe the five frequently occuring syntatic patterns in reviews that are used to extract candidate text segments.
We then discuss the use of topic models in selecting meaningful segments from the set of extracted candidates.
We also present an independent framework for evaluation of the summaries comprising segments
 regardless of the approaches.

\subsection{Extraction Patterns}
Central to our summarization system is how to extract meaningful, informative text segments out of a sentence.
We use sentence syntax to guide the extraction process by defining patterns of lexical classes for matching against
 text segments. The purpose is to extract semantically meaningful unit of text in a sentence that can be understood without extra context.  In the particular task of summarizing reviews for products and services, we want to capture units that contain sentiments toward aspects.  This type of segments is important because it expresses and formulates opinions about the entity being reviewed. 

Based on the above observation, we identify five most common extraction patterns to capture a variety of text
segments in both product and service reviews as follows. First we use POS tagger to tag all pros and cons items available in our data
 sets of restaurant and coffee maker reviews  (see Sect. 5.1). The pros and cons are relatively short and meaningful, and can therefore be suitable  representatives of the text segments that we want to generate. The resulting sequences of tags are then ranked based on their frequency.  After carefully studying the top ranked patterns we select the five most productive ones listed in Table ~\ref{tab:patterns}.
  \begin{table}
\centering
\caption{Extraction patterns and their occurrences in data sets}
\begin{tabular}{c|c|p{6.5cm}|c|c}
\textbf{no.} & \textbf{the pattern} & \textbf{example} & \textbf{restaur} & \textbf{coff makers} \\ \hline
1&nn? vb dt? rb* jj nn &  instruction booklet includes clear instruction &56468 &23210 \\ \hline
2&nn? vb rb* jj to vb & filter basket is simple to remove &4226&3770 \\ \hline
3&nn? vb rb* jj  & design is striking, tasted fresh &130853&30449\\ \hline
4&rb* jj to vb nn? & easy to clean, wide enough to insert a K-Cup &5937&5288\\ \hline
5&rb* jj nn & very good food, most expensive pod brewer &197123&69273\\ \hline
\end{tabular}
\label{tab:patterns}
\end{table}

We use the same notations as in regular expression, where the constituent parts correspond to lexical categories as specified by the PennTree bank. For simplicity, a single tag is used to represent different forms of a category; i.e., \textbf{jj} represents adjective and matches all of JJ, JJR and JJS.
Also, \textbf{nn} matches a noun phrase rather than just a single word. We further restrict that each segment must match the longest pattern.
 This means, for example, a segment matching pattern 1 in a sentence is consumed and no longer available for matching pattern 5.
 Each pattern also has its negation form easily constructed from its positive form, hence we do not show in the table.

\subsection{Selecting Informative Segments using Topic Models}
\label{sec:selection}
Candidate segments can be meaningless even if they match
the defined extraction patterns. For example, `final thought'
and `several hour' are instances of pattern 5, but they
reveal no interesting information.
Furthermore, the sheer number of text segments
matching the patterns (Table ~\ref{tab:patterns})
requires us to be selective in finding segments to include in summaries. 

We observe that informative segments often contain words that convey opinions about aspects of entities. Since the aspect-sentiment intimate interplay is modeled and learned by our joint sentiment-topic model, we propose the following filters to prune less informative segments using the output the model.
\begin{description}
\item[Baseline] No filtering, i.e., keep all matching segments.
\item[AW\quad\quad] Eliminate a segment if it does not contain one of the top $X$ most probable words of the segment's inferred aspect.
\item[SW\quad\quad] Eliminate a segment if it does not contain one of the top $Y$ most probable sentiment words of the segment's inferred sentiment and aspect.
\item[RANK\quad] Rank all segments having the same inferred sentiment and aspect in order of their probabilities and eliminate the bottom half segments.
\end{description}
It is possible to use previous joint sentiment topic models, such as ASUM \cite{Jo:asum} and JST \cite{Lin:jst}, for the filtering purpose.
Note that ASUM and JST output word distributions for each pair of sentiment and aspect; hence, \textbf{ASUM} and \textbf{JST} are in effect both sentiment classifier and filter: 
\begin{description}
\item[ASUM] Eliminate a segment if it does not contain one of the top $Z$ most probable words of the segment's inferred senti-aspect using the ASUM model.
\item[JST] Same as \textbf{ASUM} except that JST is used in placed of ASUM.
\end{description}

A complete procedure for summarization would need a sentiment classifier component for segments as sentiment-based summaries are preferred by users [10]. In addition to our model-based sentiment classifier, we introduce another
sentiment classifier based on SentiWordNet (SWN) [4], a popular lexical resource for opinion mining, using the same approach as in \cite{Fahrni:oldwine}. For convenience, we call our model-based classifier \textbf{SEN} and the SWN-based classifier \textbf{SWN}.

Various procedures for retaining quality segments can then be constructed by combining different sentiment classifiers and filters.
 For example, we may first use \textbf{SEN} to classify sentiment of a segment and then use both \textbf{AW} and \textbf{SW} to
 discard non-qualified segments. We name such procedure \textbf{SEN+AW+SW}, with the convention that the output of a 
 preceding classifier/filter is the input to the next classifier/filter whenever applicable.

\subsection{A Framework for Segment-based Summary Evaluation}
\label{sec:framework}
We now introduce a framework for automatically evaluating the extraction patterns at the levels of segment and
entity (a specific product or service). This framework is independent of the way segments are generated and therefore
can be applied to any method that uses segment as the unit of summary.

Each entity E has a candidate summary $E^{C} = \{Y | Y$ matches one of the patterns $\}$ and a reference summary $E^{R} = \{X | X$ is in the gold standard summary of E $\}$. 
For $Y \in E^{C}$ and $X \in E^{R}$, we measure the similarity of their content using precision and recall
\begin{alignat*}{2}
P(X, Y) &:= \frac{skip2(X, Y)}{\binom{|Y|}{2}}\enspace,
&\quad\quad
R(X, Y) &:= \frac{skip2(X, Y)}{\binom{|X|}{2}} 
\end{alignat*}
where $skip2(X, Y)$ is the number of skip-bigram matches between X and Y (termed ROUGE-SU in \cite{Lin:rouge}).
For a candidate segment $Y \in E_{C}$, define $R(Y) = R(X_{max},Y) \mbox{ and } P(Y) = P(X_{max}, Y)$
where $X_{max} = \arg \max_{X \in E^{R}} R(X,Y)$.

For an entity E, the average precision $P_{skip}(E)= \sum_{Y \in E^{C}} P(Y) / |E^{C}|$ and recall $R_{skip}(E) = \sum_{Y \in E^{C}} R(Y) / |E^{C}|$ tells us how similar the content of extracted segments is to a reference set of segments  on average.
We also want to assess how many portion of the reference summary is recovered and what percentage of the candidate summary is useful.
For this reason, we introduce P(E) and R(E) to measure the precision and recall for the candidate summary set $E^{C}$ of $E$:
\begin{alignat}{2}
P(E) &:= \frac{\sum_{Y \in E^{C}} \mathbf{1_{A}\{Y\}}}{|E^{C}|}, 
&\quad
R(E) &:= \frac{\sum_{X \in E^{R}} \mathbf{1_{B} \{X\} }} {|E^{R}|}
\label{eq:pr}
\end{alignat}
where $\mathbf{1_{A}\{Y\}}$ and $\mathbf{1_{B} \{X\} }$ are indicator functions;
$\mathbf{A} = \{ Y | $ $R(Y) \ge \alpha\}$ and $\mathbf{B} = \{X | \text{ } \exists Y \in \mathbf{A}$ s.t $R(X, Y) = R(Y) \}$ where $\alpha$ is a recall threshold for a candidate summary to be considered useful.

A good measure for a reference summary of an entity $E$ must be a combination of the segment-level recall (precision),
 $R_{skip}(E)$ and the entity-level recall (precision), $R(E)$. A simple combination is the average of the two, i.e, $R_{cb}(E) = (R_{skip}(E) + R(E))/2$ and $P_{cb}(E) = (P_{skip}(E) + P(E))/2$.

Since we typically work with data that contains a large set of review entities, it is convenient to report the results using the following summarization statistics: 
\begin{alignat*}{3}
P_{s} &= \frac{\sum_{i = 1}^{N}\sum_{Y \in E^{C}_{i}}P(Y)}{\sum_{i = 1}^{N}|E_{i}^{C}|}\enspace,
&\quad
P_{e} &= \sum_{i = 1}^{N} P(E_{i}) / N \enspace,
&\quad
R_{e} &= \sum_{i = 1}^{N} R(E_{i}) / N \enspace,\\
R_{s} &= \frac{\sum_{i = 1}^{N}\sum_{Y \in E^{C}_{i}}R(Y)}{\sum_{i = 1}^{N}|E_{i}^{C}|}\enspace,
&\quad
P &= \sum_{i = 1}^{N} P_{cb}(E_{i}) / N \enspace,
&\quad
R &= \sum_{i = 1}^{N} R_{cb}(E_{i}) / N\enspace. 
\end{alignat*}

\section{Experiments}
We experimented using reviews of coffee makers as representative for the product domain and reviews of restaurants
as representative for the service domain. We describe our data sets and experimental set-ups in ~\ref{sec:data}.
 In ~\ref{sec:topicsandpolarities} we give example of the topics and sentiment words learned by the model. We
analyze the effectiveness of extraction patterns in ~\ref{sec:patternsevaluation} and compare the performance of different sentiment classifiers and segments filters in ~\ref{sec:selectionevaluation}.

\subsection{Data Sets and Experimental Set-ups}
\label{sec:data}
Our data sets consist of restaurant reviews and coffee maker reviews. For each review, we collected its free-format text content and its pros and cons lists if available.
\begin{itemize}
\item{\textbf{RESTAURANTS} 50,000 reviews of  5,532 restaurants collected from Citysearch New York. This data is provided by Ganu, et al. \cite{Ganu:ursa}}.
\item{\textbf{COFFEEMAKERS} 23,411 reviews of 534 coffee makers collected from epinions.com.}
\end{itemize}

Our first step is to fit the joint sentiment topic model to each data set.
Data is pre-processed as in other standard topic models, in which sentences are tokenized by the punctuations: `.', `!', and `?'. The hyperparameters are set as $\alpha$ = 0.1, $\beta$ = 0.01, 
$\gamma$ = 0.1 for both positive and negative sentiment; the number of aspects is 7 for both corpora. 

We incorporated prior sentiment information into the model using sentiment seed words in analogy to [9].
After running the sampler for a burnin period of 500 iterations, we interleaved it with the optimizer,
optimizing over $y_{ki}$ and $y_{ji}$ every $100^{th}$ step of sampling. We
trained the model in 2000 iterations for both data sets and
used the last sample in the chain in all of our experiments.

 In the segments selection step, the maximum number of words in
a sequence is set to 7 and the number of top words for \textbf{AW}, \textbf{SW}, \textbf{ASUM}, and \textbf{JST} is set to 200, 100, 300, and 300, respectively. We used a value of 0.25 for the recall threshold
$\alpha$ in Eq.~\ref{eq:pr}. All parameters were set empirically after
many experiments.

In order to evaluate the quality of segments and summaries using the framework in ~\ref{sec:framework},
 a reference summary must be obtained for each review entity. We aggregate the pros written by all reviewers for an entity as its pros gold standard and similarly for its cons standard (duplicated entries are removed). To construct an entity's candidate summaries, the procedures in ~\ref{sec:selection} are applied to the segments extracted from all of its reviews. The sentiment classifier in a procedure partitions the
entity's segments into a positive candidate summary and a negative candidate summary. The candidates are evaluated against their counterpart reference summaries independently. 
\begin{table}
\centering
\caption{Example inferred topics (restaurants: row 1-3, coffee makers: row 4-6)}
\begin{tabular}{p{4.4cm}|p{4.45cm}|p{4.4cm}}
\textbf{top aspect words} & \textbf{top positive words} & \textbf{top negative words} \\ \hline
sauc, chicken, chees, salad, shrimp, soup, fri, potato, rice & good, delici, best, great, fresh, love, perfect, excel, amaz, tasti & dri, disappoint, tasteless, cold, soggi, bad, fri, rare, medium\\ \hline
music, place, bar, decor, room, table, wall, seat, atmosphere & great, nice,  good, love, beauti, enjoy, romant, perfect, friend, & loud, noisi, bad, littl, small, crowd, dark, expens, back\\ \hline
wait, table, waiter, seat, minut, reserv, order, ask, told, manag & friend, nice, worth, great, attent, prompt, long, enjoy, quick & rude, bad, wrong, final, empti, horribl, terribl, poor, worst \\ \hline
coffe, bean, cup, grind, ground, brew, grinder, espresso, tast & good, like, fresh, great, best, hot, strong, fine, french, perfect  & weak, bad, disappoint, wast, grind, wrong, unfortun, bitter \\ \hline
filter, clean, basket, water, paper, rins, dishwash, gold, use & easi, clean, perman, like, remov, easili, good, recommend, safe & difficult, clean, wet, bad, imposs, perman, wast, not easi \\ \hline
servic, game, custom, warranti, repair, ship, product, send, call & good, new, back, great, free, thank, well, happi, local, origin, & back, disappoint, poor, bad, wrong, negative, defect, sorri 
\end{tabular}
\label{tab:topics}
\end{table}

\subsection{Topics and Polarities of Sentiment Words Learned by the Model}
\label{sec:topicsandpolarities}
Example of topics inferred by the model is given in Table
~\ref{tab:topics}. Each topic has three
 distributions where one distribution (first column) consists 
descriptive words about the aspect and two distributions
(remaining columns) consist evaluative words directing the aspect.
Except the common sentiment words such as \textit{good}, \textit{great}, \textit{bad}, \textit{wrong}
that are associated with most aspects due to their frequent usage, positive and negative sentiment lexicons look highly related to their corresponding aspects. For example, the model discovers that people
are more likely to praise the food with \textit{delicious}, \textit{best}, \textit{fresh}, and \textit{tasty}
 and disapprove food that is \textit{dry}, \textit{tasteless}, \textit{cold} or \textit{soggy}. Such results can be very helpful
for the exploratory purpose of understanding what aspects reviewers care and comment about.

Table ~\ref{tab:lexicons} demonstrates the effectiveness of our model in learning the polarities of domain-specific sentiment lexicons 
 (the seed words used for bootstrapping are excluded). To verify this claim we compare with the \textbf{SWN} classifier described in ~\ref{sec:selection} in a classification task for noun phrases. SWN leverages synsets in WordNet and so, in some sense, it captures the context-dependent sentiment of a word.
We used a set of 929 positive and 236 negative noun phrases obtained from an external set of restaurant reviews
in [5]. All phrases are unique and manually annotated with their true sentiments. Our classifier outperforms \textbf{SWN} in classification accuracy for both the positive (90.1\% vs. 83.4\%) and and negative (74.6\% vs. 66.1\%) categories. This shows that our model is quite accurate in  assigning sentiment score to domain-specific lexicons compared to the more general propagation approach in SWN.
\begin{table}
\centering
\caption{Selected lexicons with their sentiment polarities}
\begin{tabular}{p{12cm}} \hline
\textbf{positive lexicons in restaurant reviews} \\ \hline
knowledgeable, helpful, unique, courteous, prompt, cozy, terrific, wonderful, affordable, superb, warm, impeccable, outstanding, elegant, consistent, fabulous, charming \\ \hline
\textbf{negative lexicons in restaurant reviews} \\ \hline
tasteless, mediocre, bland, inedible, dry, ridiculous, lousy, overpriced, flavorless, average, unacceptable, obnoxious, soggy, bare, bore, tough, unfriendly, horrendous, stale \\ \hline
\textbf{positive lexicons in coffee maker reviews} \\ \hline 
simple, ready, fresh, correct, removable, automatic, impressive, stainless, large, free, light, strong, rich, reasonable, amazing, fast, clear, wonderful, delicious, quick, sturdy\\ \hline
\textbf{negative lexicons in coffee maker reviews} \\ \hline 
difficult, impossible, inferior, loud, lousy, dull, defective, stupid, sticky, dirty, faulty, uneven, weak, noisy, stiff, frustrating, dissatisfied, smelly, unclear, erratic, leak, slow\\ \hline
\end{tabular}
\label{tab:lexicons}
\end{table}
\subsection{Evaluation of Extraction Patterns}
\label{sec:patternsevaluation}
We now analyze how different extraction patterns behave when applied to the service domain and the product domain
(Table ~\ref{tab:individual1} and ~\ref{tab:individual2}). We use \textbf{AW+SEN+SW} procedure because it produced the best result among all methods.
For reviews of restaurants, pattern 3 and 5 are the most
productive with superior average precision and recall at both
segment-level and entity-level compared to the rest. They account for more than
half of an entity's pros and cons reference. This is probably due to the prevalence of sentences such as `the service was good' in restaurant reviews. The result is consistent with the current literature where
 adjectives and nouns are commonly used to detect sentiments and aspects in reviews for services.
It is worth noticing that extracting any thing other than adjective-noun pairs may degrade the quality of summarization as 
the scores for pattern 2 and 4 are overwhelmingly low. 
\begin{table}
\centering
\caption{Comparison of extraction patterns for services.}
\begin{tabular}{c|l|l|l|l|l|l|l|l|} \cline{2-9}
\label{tab:individual1}
&\multicolumn{4}{|c|}{pros} & \multicolumn{4}{|c|}{cons}\\ \hline
patt& $\mathrm{P_{s}}$ & $\mathrm{R_{s}}$ & $\mathrm{P_{e}}$ & $\mathrm{R_{e}}$ & $\mathrm{P_{s}}$ & $\mathrm{R_{s}}$ & $\mathrm{P_{e}}$ & $\mathrm{R_{e}}$\\ \hline
1&20.2	&51.1	&74.4	&30.4  &14.6	&\textbf{39.6}	&55.7	&14.8\\ \hline
2&23.7	&26.4	&54.6	&1.9  &7.9	&22.4	&63.6	&0.5 \\ \hline
3&31.8	&\textbf{65.7}	&83.3	&\textbf{40.6}  &26.4	&\textbf{57.6}	&70.1	&\textbf{21.5} \\ \hline
4&21.3	&28.8	&56.1	&2.6  &6.3 &15.5	&35.7	&0.37 \\ \hline
5&25.9	&\textbf{53.5}	&72.7	&\textbf{49.7}  &18.1	&37.8	&52.7	&\textbf{31.2} \\ \hline
\end{tabular}
\end{table}

The behaviors of extraction patterns are trickier for the product domain as can be seen in Table ~\ref{tab:individual2}.
 There is no dominating pattern in terms of high precision and recall
at both segment and entity level. In particular, pattern 3
and 5 still recover a large portion of an entity's reference
summary; however, the average quality of their matching
segments ($R_{s}$) is the lowest among all patterns. Pattern 2
and 4 perform badly when used with the service domain but
are more useful in the product domain, producing
the highest quality segments ($R_{s} = 66.3 \text{ and } 69.4$ for positive; $R_{s} = 48.6 \text{ and } 43.8$ for negative).
Although they do not appear as frequently in reviews as other patterns, they
tend to carry more meaning in their words that it is hard to
ignore them. Hence, all five patterns can contribute to the
extraction of informative segments for summarization. This
shows that doing summarization for products is harder than
for services; and, care should be exercised when generalizing
results from one domain to the other.
\begin{table}
\centering
\caption{Comparison of extraction patterns for products.}
\begin{tabular}{c|l|l|l|l|l|l|l|l|} \cline{2-9}
\label{tab:individual2}
&\multicolumn{4}{|c|}{pros} & \multicolumn{4}{|c|}{cons}\\ \hline
patt& $\mathrm{P_{s}}$ & $\mathrm{R_{s}}$ & $\mathrm{P_{e}}$ & $\mathrm{R_{e}}$ & $\mathrm{P_{s}}$ & $\mathrm{R_{s}}$ & $\mathrm{P_{e}}$ & $\mathrm{R_{e}}$\\ \hline
1&22.1	&59.6 &70.8	&\textbf{30.1}  &19.5 &\textbf{44.2} &59.7 &\textbf{17.5} \\ \hline
2&45.5	&\textbf{66.3}	&78.7	&6.6    &30.2	&\textbf{48.6}		&64.3	&2.9\\ \hline
3&33.2	&54.4	&65.9	&26.7   &27.0 	 &37.7		&51.1	&15.0\\ \hline
4&52.7	&\textbf{69.4}	&78.5	&8.6   &30.9	&43.8 &61.3	&3.5\\ \hline
5&26.0	&50.8 &59.6	&\textbf{38.9}   &26.2	&40.8	&56.3	&\textbf{25.0}\\ \hline
\end{tabular}
\end{table}

\subsection{Evaluation of Sentiment Classifiers and Segment Filters}
\label{sec:selectionevaluation}
Results in the previous section suggest to use different syntactic patterns
 for summarization of the service and product
domains. We used patterns 1, 3, and 5 for services and all
patterns for products in all of our experiments in this section.

We applied seven different procedures for selecting candidate segments to compare the effects of 2 sentiment classifiers
 (\textbf{SEN} and \textbf{SWN}) and 5 filters (\textbf{AW}, \textbf{SW}, \textbf{RANK}, \textbf{ASUM}, and \textbf{JST}).
 The results are depicted in Table ~\ref{tab:services} and ~\ref{tab:products}. The good overall performance of the \textbf{Baseline+SWN} procedure in both domains indicates that the proposed patterns extract good segments for summarization. 
\begin{table}
\centering
\caption{Comparison of classifiers and filters for services.}
\begin{tabular}{c|l|l|l|l|l|l|l|l|l|l|l|l|} \cline{2-13}
\label{tab:services}
&\multicolumn{6}{|c|}{pros} & \multicolumn{6}{|c|}{cons}\\ \hline
procedure & $\mathrm{P_{s}}$ & $\mathrm{R_{s}}$ & $\mathrm{P_{e}}$ & $\mathrm{R_{e}}$ & P & R & $\mathrm{P_{s}}$ & $\mathrm{R_{s}}$ & $\mathrm{P_{e}}$ & $\mathrm{R_{e}}$ & P & R\\ \hline
Baseline+SWN &24.4	&48.6	&64.8	&65.4	&44.5	&56.7	  & 17.6	&29.0	&42.1	&45.0	&29.7	&36.8\\ \hline
AW+SWN  &24.8	&55.5	&71.4	&60.4	&47.9	&57.6   &19.0	&33.8	&47.6	&37.8	&33.1	&35.5\\ \hline
AW+SEN  &23.8	&52.3 &67.5	&62.8	&45.5	&57.2 &21.3	&47.6 &61.5	&36.9	&41.3	&42.1\\ \hline
AW+SEN+RANK  &\textbf{29.3}	&52.6 &66.7	&44.5	&47.8	 &48.2 &\textbf{26.0} &46.8 &60.4	&23.1	&43.0	&34.8 \\ \hline
AW+SEN+SW & 25.8 &	\textbf{56.4} & \textbf{73.3} &	59.9 &	\textbf{49.3} &	\textbf{57.8} & 25.4&	\textbf{58.2}& \textbf{73.5}&	30.2&	\textbf{49.4}&	\textbf{44.0} \\ \hline
ASUM& 27.4&	49.7& 66.7&	\textbf{65.8}&	46.8&	57.4& 22.0&	33.5& 47.0& \textbf{47.4}&	34.3&	40.2 \\ \hline
JST &24.6	&44.7  &59.8	&60.4	&42.0	&51.9 &21.3	&37.3 &49.4 &52.3	&35.2	&43.9 \\ \hline
\end{tabular}
\centering
\caption{Comparison of classifiers and filters for products.}
\begin{tabular}{c|l|l|l|l|l|l|l|l|l|l|l|l|} \cline{2-13}
\label{tab:products}
&\multicolumn{6}{|c|}{pros} & \multicolumn{6}{|c|}{cons}\\ \hline
procedure & $\mathrm{P_{s}}$ & $\mathrm{R_{s}}$ & $\mathrm{P_{e}}$ & $\mathrm{R_{e}}$ & P & R & $\mathrm{P_{s}}$ & $\mathrm{R_{s}}$ & $\mathrm{P_{e}}$ & $\mathrm{R_{e}}$ & P & R\\ \hline
Baseline+SWN&23.1	&49.2	&60.9	&44.8	&41.1	&43.9	&23.9	&34.5	&47.3	&\textbf{32.6}	&34.6	&30.7 \\ \hline
AW+SWN&21.8	&52.6	&65.4	&42.7	&42.8	&44.7	&23.4	&39.0	&52.7	&30.8	&37.6	&31.6 \\ \hline
AW+SEN &23.9	&56.3	&68.5	&47.0	&45.3	&48.2	&26.1	&47.5	&62.9	&24.7	&43.5	&\textbf{32.1} \\ \hline
AW+SEN+RANK&\textbf{29.7}	&51.2	&61.0	&32.4	&44.0	&38.4	&\textbf{32.6}	&37.3 &48.9	&21.2	&39.4	&25.8 \\ \hline
AW+SEN+SW &25.5	&\textbf{59.8} &\textbf{71.4}	&43.8	&\textbf{47.4}	&48.2 &24.3	&\textbf{51.4} &\textbf{65.4}	&16.7	&\textbf{44.2}	&29.4 \\ \hline
ASUM &25.3	&54.8 &68.2	&\textbf{49.3}	&45.7	&\textbf{48.8} & 26.9	&37.8& 49.0	&28.7&	36.7	&29.8  \\ \hline
JST &27.1 &	52.0& 64.3	&44.9	&44.7	&45.2 &25.9	&35.7 &46.4&	28.8	&34.5	&28.5 \\ \hline
\end{tabular}
\end{table}

Comparing \textbf{AW+SWN} and \textbf{AW+SEN}, we see that \textbf{SEN} is better than \textbf{SWN} at sentiment classification.
 This result agrees with previous section, again confirming the effectiveness of our model in learning sentiments
 of domain-specific lexicons. \textbf{AW+SWN} performs better than \textbf{Baseline+SWN},
 suggesting that top aspect words can be used to identify more informative segments.
 The best procedure is \textbf{AW+SEN+SW}, which tops all other procedures especially in the cons case.
 It favors segments that contain common aspect-related words and its associated sentiment lexicons,
 which are likely to be predominant in the pros and cons lists.
ASUM has similar modelling assumption as ours and so it also produces relatively good results. 
 However, the ability of our model to optimize sentiment polarities creates the improvement in performance.
 \textbf{JST} is even inferior to \textbf{Baseline+SWN} for half of the cases. This is not surprising given
 that the JST model is not intended for sentiment lexicons discovery; in contrast, it requires a large list of sentiment
 seed words to function well. Finally, \textbf{AW+SEN+RANK} always has highest precision for segments but many segments
 are eliminated and that hurts its performance. 

\section{Qualitative Evaluation}
\label{sec:qualitative}
In this section we complement our results in previous section by
 qualitatively evaluating the quality of extracted segments
 with a user study and present example of summaries generated by our approach.
 
\subsection{Quality of Extracted Segments}
We carried out an user study with 130 workers from the
Amazon Mechanical Turk service. We randomly selected 123
short passages each has 4 to 6 sentences from reviews of
 coffee makers. The user's task is to read a passage and rate each text item as `very useful',
`useful', `somewhat useful', or `useless' with reference to the passage.  We included two types of items for each passage: segments
extracted by our approach using the \textbf{AW} filter and adjective-noun phrases extracted using tagging and term-frequency as in \cite{Yatani:spotlight}.
Each user performs 6 tasks in which half of them are repetitions
of others, thereby allowing us to detect users that give inconsistent ratings. We discarded users who completed their tasks in less than 90 seconds or rated half of the items inconsistently. Of the remaining 90 qualified users, 13 have not used coffee makers before whereas 60 have used for more than two months.

In total there were 358 unique segments, each rated 5.3 times and 470 unique word pairs, each rated 5.7 times. We converted the ratings into a numeric scale from 4 to 1 with 4 being 'very useful' and 1 being 'useless'. On average, users rated the segments extracted by our method as 3.01 compared to 2.53 for the adjective-noun phrases. The higher rating is not merely due to segments having more words, as we observed that users typically give an adjective-noun word pair a same or higher rating than
 a segment if the two carry the same message. For example, `carafe stays hot' and `hot carafe' are same but the former
has a rating of 2.7 whereas the latter has a rating of 3.1. Therefore the higher average rating for segments is
a strong evidence that they convey more valuable information than adjective-noun word pairs.

 Table ~\ref{tab:highlyrated} elaborates further on this evidence by showing
example of the segments and phrases rated as very useful by users.
 As can be seen, the segments are quite complete semantically
 whereas the phrases can be rather short in their meaning, which may require interpretation from users.
\begin{table}
\centering
\caption{Example of highly rated segments and phrases}
\begin{tabular}{p{12cm}} \hline
\textbf{segment} \\ \hline \hline
very easy instruction, almost completely unscrewed to pour, buttons are easy to press, cup is always fresh, coffee pot is very hard to take, closing is easy, makes really awful tasting coffee, feature works fine, machine brews a great cup every time, machine is very simple to use, machine is programmable, carafe is dripless  \\ \hline
\textbf{adjective-noun phrase} \\ \hline \hline
affordable maker, better tasting, correct time, filtered water, finished quality, difficult place, fresh tasting, good customer, good tasting, new recipe, removable basket, optimal temperature, cheap use,  easy closing, darker flavor, hot cup, great pot  \\ \hline
\end{tabular}
\label{tab:highlyrated}
\end{table}
 \subsection{Example Summaries}
Below we show examples of a restaurant review and a coffee maker review together
 with the segments extracted as their summaries.

\textbf{Review of restaurant}: \textit{The space is small but cozy, and the staff is friendly and knowledgeable. There was some great music playing, which kind of made me feel like I was on vacation some place far away from Astoria. There are a lot of really great vegetarian options, as well as several authentic Turkish dishes.  If you're still wasting time reading this review, stop now and head straight for Mundo.  Your stomach could already be filled with tons of deliciousness.}

\textbf{Summary}: staff is friendly, space is small, some great music playing , several authentic Turkish dishes, really great vegetarian options.

\textbf{Review of coffee maker}: \textit{I bought this machine about a week ago. I did not know which machine in the store to get, but the sales clerk helped me make the decision to buy this one. It is incredibly simple to use and the espresso is great. The crema is perfect too. My latte's rival those in coffee houses and I am saving a ton of money. The "capsules" must be ordered from the Nespresso website, but they are usually at your door in 48 hours via UPS...}

\textbf{Summary}: incredibly simple to use, espresso is great, crema is perfect.

In both cases the summaries express the gist of each review
relatively well. Looking at the sentence where a segment is
extracted from, it can be seen that the segment conveys the
main talking point of the sentence. Additionally,
 each segment does express an opinion about some aspect of the coffee
maker or the restaurant. Recall that our key assumption in
modeling reviews is that each sentence has a sentiment and
an aspect. Therefore extracting segments the way we propose
 is likely to capture the main content of a sentence.

\section{Conclusions}
In this paper we have describe a framework for extracting and selecting informative
segments for review summarization of products and services. We extract candidate segments by 
matching against variable-length syntactic patterns and select the segments that contain top sentiment
and aspect words learned by topic models. We proposed a new joint sentiment topic model that learns the
polarity of aspect dependent sentiment lexicons. Qualitative and quantitative experiments verify that our model outperforms previous approaches in improving the quality of  the extracted segments as well as the generated summaries. 
\bibliographystyle{splncs03}

\end{document}